\documentclass[preprint,showpacs,preprintnumbers,amsmath,amssymb]{revtex4}

\usepackage{graphicx}
\usepackage{dcolumn}
\usepackage{bm}
 
\begin{document}
\draft
\preprint{}

\title{Instability of quasi-liquid on the edges and vertices of 
       snow crystals }
\author{Kanako T. Sato}
\email{kana@jiro.c.u-tokyo.ac.jp}
\affiliation
{Department of Pure and Applied Sciences,University of Tokyo, 
Komaba 3-8-1, Meguro-ku, Tokyo 153-8902, Japan}

\date{\today}

\begin{abstract}
In this paper, 
we show theoretically that 
there exists quasi-liquid on the edges and vertices of snow crystals 
between $-4^{\circ}$C and $-22^{\circ}$ C, 
while the faces 
$(0001)$ and $(10{\bar 1}0)$ have no quasi-liquid layers.
Investigating the macroscopic theory of quasi-liquid 
and applying to the edges and vertices of the crystal, 
we find that the quasi-liquid 
becomes unstable above the critical supersaturation point, 
which is above the water saturation point.
The thickness of this unstable quasi-liquid layer continues 
growing indefinitely. 
We interpret this behavior as corresponding to continuous production 
and overflow onto neighboring faces in a real system. 
We hypothesize that  
the unstable growth of snow crystals 
originates from the edges and vertices, and it 
is due to the overflow of quasi-liquid 
from the edges and  vertices onto the neighboring faces,  
which are rough and lack quasi-liquid. 
Our hypothesis 
accounts for the qualitative behavior of the relations 
 between the morphological instability  and the water saturation 
in the snow phase diagram.
\end{abstract}


\maketitle

\section{\label{sec: 1}Introduction}

\begin{figure*}
\includegraphics[width=12cm]{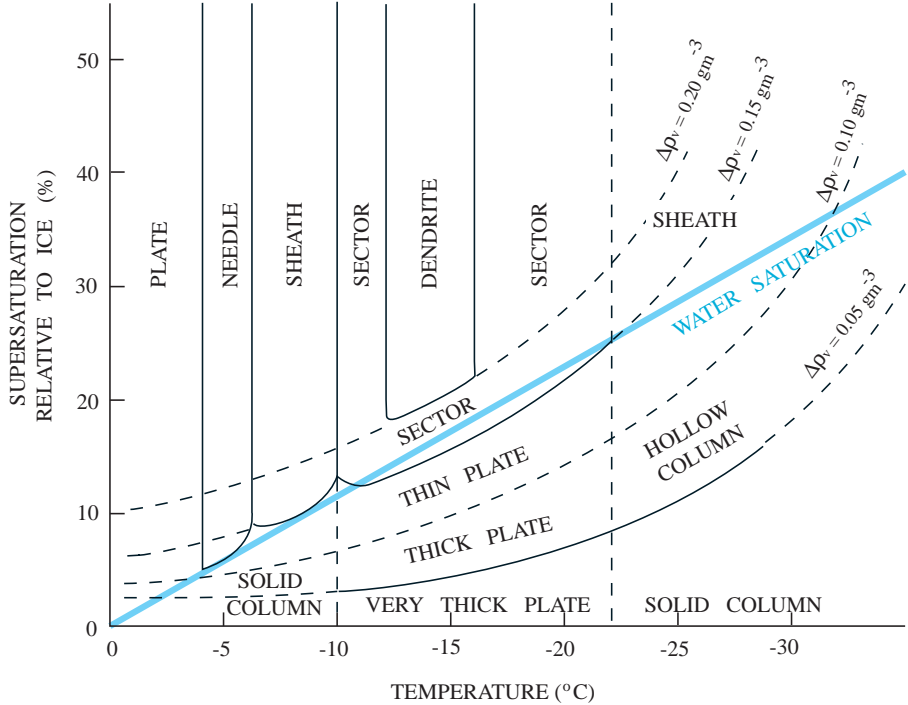}
\caption{\label{fig: snowdiagram}
Variation of ice crystal form with temperature and
 supersaturation. Here $\Delta \rho_{\rm v}$ is the vapor density excess.  
(Based on laboratory observations \cite{Aufm}.)}   
\end{figure*}

The snow crystal is one of the most beautiful things in nature, 
and it has attracted human interest since long ago. 
Its basic form is a hexagonal prism bounded by two basal $(0001)$ 
and six prism faces $(10{\bar 1} 0)$.  
Nakaya \cite{Nakaya} was the first to successfully produce 
snow crystals in the laboratory, 
and he investigated the relations between 
growth forms and experimental conditions 
(i.e., temperature and supersaturation relative to ice). 
Since his work, many experimental studies 
have been carried out \cite{Aufm}. 
The snow phase diagrams obtained in these studies 
exhibit two important features (See Fig.\ref{fig: snowdiagram}).
One is that three transitions occur in the basic crystal form: 
at temperatures near 
$-4^{\circ}{\rm C}$ (plates to columns), 
$-10^{\circ}{\rm C}$ (columns to plates), and  
$-22^{\circ}{\rm C}$ (plates to columns). 
These transitions 
correspond to changes in the growth 
rates of the faces $(0001)$ and the faces $(10{\bar 1}0)$ 
with temperature \cite{Lamb}. 
The second important point is that 
a morphological instability arises  
when the supersaturation relative to ice becomes sufficiently large. 
Especially above the water saturation point 
(saturation relative to supercooled water), 
the crystal growth originates from the edges and the vertices, 
and as a result, 
needles and sheaths are produced 
between $-4^{\circ}{\rm C}$ and $ -10^{\circ}{\rm C}$,  
and dendrites and sectors are produced between $-10^{\circ}{\rm C}$ and 
$-22^{\circ}{\rm C}$. 

The basic growth form of snow crystals is 
determined by the most slowly growing faces. 
The basal $(0001)$ and prism $(10{\bar 1}0)$ faces grow slowly 
and become the bounding faces of the crystal. 
Assuming that the faces grow through 
the spreading motions of two-dimensional nuclei    
\cite{BCF}, 
and taking into account the vapor diffusion field surrounding the crystal, 
Frank proposed a possible explanation of the dendritic growth of snow
crystals \cite{Frank}. 
Based on the same idea, Yokoyama and Kuroda \cite{YK} simulated the
growth of an infinitely long hexagonal column 
(i.e. a two-dimensional snow crystal), 
and they succeeded in showing 
that six primary branches are produced from the vertices of the column. 
The instability that results in the production of these branches
is due to the nonuniformity of the 
vapor density surrounding the polyhedron crystal. 
Their simulation, however, also showed that 
when secondary branches are produced,
the size of the primary branches becomes comparable with
that of the original hexagonal column.  
This is not a realistic modeling 
of the dendrites in actual snow crystals.
Furthermore, 
this theory cannot account for the snow phase diagram discussed above.   
It is thus apparent that 
we  must take into account other, yet unknown effects to explain
the dendritical instability.

The three experimentally observed transitions  
in the basic crystal form at low supersaturation 
were accounted by the theory of Kuroda and Lacmann \cite{KurodaLacmann}.
This theory is based on the well-known fact that 
the surface of ice existing just below 
$0^{\circ}{\rm C}$ is coated with a thin liquid like layer
(quasi-liquid layer). 
This quasi-liquid layer has been observed using a variety of 
experimental techniques \cite{Kvli,Mizuno,Beag,YMT,GJ,Bil,Elbaum}.  
Theoretically, 
Weyl \cite{Weyl} gave qualitative arguments for its existence, 
and Fletcher \cite{Flet} developed Weyl's arguments
into a quantitative form.   
The starting point of Kuroda and Lacmann is a phenomenological
model of the quasi-liquid layer proposed by Lacmann and Stranski
\cite{LS1}.
On the basis of this model, 
Kuroda and Lacmann predicted that 
there exists quasi-liquid on the face $(10{\bar 1}0)$ at lower 
temperatures than on the face $(0001)$.     
Recently, 
it has been observed in both experiments \cite{YMT, Bil} and 
molecular dynamics simulations \cite{Nada} 
that there is a difference in behaviour of the quasi-liquid layer 
on basal and prism faces. 
Kuroda and Lacmann hypothesized  
that the surface structure of ice changes with decreasing 
temperature from a surface covered with a quasi-liquid 
layer to a rough surface without a quasi-liquid layer 
and finally to a smooth surface.
According to their description,  
the changes in crystal form are due to the anisotropy  
in these surface structural transitions 
between basal and prism faces of ice.

In this paper, we investigate the interaction 
between the quasi-liquid and the edges and vertices of snow crystals, 
whose growth originates from the edges and the vertices 
from $-4^{\circ}$C to $-22^{\circ}$C and above the water saturation point.
In Section 2, we present a macroscopic model of a quasi-liquid layer 
and the explanation of the crystal form transitions proposed 
by Kuroda and Lacmann. 
In Section 3, we give an argument asserting that quasi-liquids 
remain on the edges and  vertices in the temperature region between 
$-4^{\circ}$ C and $-22^{\circ}$C,   
where the faces $(0001)$ and  $(10{\bar 1}0)$ have no quasi-liquid layer.  
In Section 4, to examine curvature effects on the quasi-liquid,
we investigate a model of an ice particle covered with
a quasi-liquid layer in a vapor environment.
In Section 5, we construct a macroscopic model describing a snow crystal 
whose six prism faces are without quasi-liquid 
but whose edges are covered with  quasi-liquid.
Using this model, we show that a novel instability arises on the edges
and the vertices above the water saturation point 
due to the presence of the quasi-liquid.
In Section 6, we present a new description of the relation 
between  morphological instability and  water saturation 
 in the snow phase diagram. 
In Section 7, we give a summary.

\section{Quasi-liquid layer on a flat surface}
\label{sec: 2}

A quasi-liquid layer can exist in a stable state 
because its presence reduces the surface free energy of the system.
The energetic advantage of surface melting is 
represented by $\Delta \sigma$ \cite{LS1}, defined as 
\begin{eqnarray}
 \Delta \sigma =\sigma_{\rm vs}-\sigma_{\rm s \ell}-\sigma_{\rm \ell v},  
\end{eqnarray}
where $\sigma_{\rm vs}$ is 
the surface energy per unit area of a vapor-solid interface,
$\sigma_{\rm s \ell}$ that of a solid-liquid one, 
and  $\sigma_{\rm \ell v}$ that of a liquid-vapor one. 
This parameter is positive for a system of water and ice, 
and thus in this case the existence of the quasi-liquid 
lowers the surface energy. 
The free energy per unit area of a 
quasi-liquid layer of thickness $\delta$  is given by
\cite{LS1, KurodaLacmann}
\begin{equation}
\Delta G_{\rm plane}(\delta)=\sigma_{\rm \ell v}+\sigma_{\rm s \ell}
+\Delta \sigma W(\delta)
+\frac{\delta}{V_{\rm q}}(\mu_{\ell}-\mu_{\rm s}), 
\label{eq: freeflat}
\end{equation}
where $V_{\rm q}$ is the molecular volume in the quasi-liquid layer and 
$\mu_{\ell}$ and $ \mu_{\rm s}$ are 
 the chemical potentials per molecule for the bulk liquid and the bulk 
solid, respectively.  
Here the function $W(\delta)$ must satisfy the conditions 
$W(0)=1$ and $W(\infty)=0$, which result from the conditions   
$\Delta G_{\rm plane}(0)=\sigma_{\rm vs}$ and 
$\Delta G_{\rm plane}(\infty)
=\sigma_{\rm \ell v}+\sigma_{\rm s\ell}
+\delta(\mu_{\rm \ell}-\mu_{\rm s})/V_{\rm q}|_{\delta=\infty}$.
To this time, two types of  $W(\delta)$ have been used 
(see, for example, Ref.\cite{DasFuWet}).  
One is short range, $W_{\rm S}(\delta)={\rm exp}(-\delta/A)$, and 
the other is long range, $W_{\rm L}(\delta)=(1+\delta/A)^{-n}$ 
\cite{KurodaLacmann}, where
$n$ is a positive integer
($n=2$ for the Van der Waals forces \cite{KurodaLacmann}) and 
$A$ is a parameter corresponding to 
the characteristic interaction length of the molecule in the 
quasi-liquid. 
Applying the minimization condition 
$\partial\Delta G_{\rm plane}/\partial\delta=0$, 
we find that   
the equilibrium thickness of the quasi-liquid layer is
\begin{eqnarray}
&& \delta_{\rm eq}^{\rm S}=-A {\rm ln} \left \lgroup 
 \frac{A Q_{\rm m} (T_{\rm m}-T)}
{\Delta \sigma V_{\rm q} T_{\rm m}}\right \rgroup,
\label{eq: Deqflat1}
\end{eqnarray}
for $W(\delta)=W_{\rm S}(\delta)$, and 
\begin{eqnarray}
\delta_{\rm eq}^{\rm L} = -A
+\left\lgroup nA^n \Delta \sigma V_{\rm q} 
\frac{T_{\rm m}}{Q_{\rm m} (T_{\rm m}-T)}\right \rgroup ^{\frac{1}{n+1}},
\label{eq: Deqflat2}
\end{eqnarray}
for $W(\delta)=W_{\rm L}(\delta)$, 
where $Q_{\rm m}$ is the energy of melting per molecule, 
$T$ is the absolute temperature, 
and $T_{\rm m}$ is the melting temperature.  
Here, we have used the equation 
$ \mu_{\ell} -\mu_{\rm s}=Q_{\rm m} (T_{\rm m}-T)/T_{\rm m}$, 
which is derived from the Gibbs-Duhem relation 
assuming constant pressure and small $T_{\rm m}-T$. 
As seen in Eqs.(\ref{eq: Deqflat1}) and (\ref{eq: Deqflat2}), 
the thickness of the quasi-liquid layer decreases monotonically with 
falling temperature.
The above equations should be considered valid 
only until the thickness becomes on the order of a few monolayers, 
but beyond this point, we can consider the layer to have vanished. 
At this temperature, the surface strongly adsorbs  
${\rm H}_2{\rm O}$ molecules, 
and it thus become rough at the molecular level. 
The number of adsorbed molecules decreases with falling temperature, 
until eventually it vanishes, and 
the surface is thus smooth.

The parameter $\Delta\sigma$ 
can be approximated in terms of the number density of the broken bonds per
unit area $\rho $ as 
\begin{eqnarray}
 \Delta \sigma =
 \frac{1}{2}\rho(\frac{Q_{\rm s}}{N_{\rm b}}-\frac{Q_{\rm m}}{N_{\rm b}})
- \sigma_{\rm \ell v}, 
\label{eq: Dsigma}
\end{eqnarray}
where $Q_{\rm s}$ is the  
energy of sublimation per molecule and 
$N_{\rm b}$ is the number of the bonds per molecule in the crystal 
\cite{Cloud}. 
For snow crystals,  
we know that  $\rho_{(10{\bar 1}0)}>\rho_{(0001)}$, 
from consideration of the surface 
molecular structures of ice, 
and $Q_{\rm s}> Q_{\rm m}$, 
from experimental tables \cite{Cloud}. 
Thus $\Delta \sigma_{(10{\bar 1}0)}>\Delta \sigma_{(0001)}$.
According to Eqs.(\ref{eq: Deqflat1}) and (\ref{eq: Deqflat2}), 
this implies that the
quasi-liquid layer on a face $(10{\bar 1}0)$ remains 
at lower temperature than that on a face $(0001)$.
Thus snow crystals have the following four
different surface structures \cite{KurodaLacmann}: 
(I) both the faces  $(10\bar{1}0)$ and $(0001)$ are covered with a 
quasi-liquid layer; 
(II) the face  $(10\bar{1}0)$ is covered with a quasi-liquid layer, 
while the face $(0001)$ has no quasi-liquid layer and is rough at 
the molecular level; 
(III)  neither has a quasi-liquid layer, but $(10\bar{1}0)$ is rough, 
while $(0001)$ is smooth at the molecular level;
(IV) neither has a quasi-liquid layer, 
and both are smooth at the molecular level. 
These structural differences result in differences  
in the growth speeds of these faces, 
since the mechanisms governing the growth depend on the surface structures.
Assuming that 
the first surface structure transforms into the second at $-4^{\circ}$C, 
the second transforms into the third at $-10^{\circ}$C, 
and the third transforms into the fourth at $-22^{\circ}$C,
Kuroda and Lacmann succeeded in describing  
the three crystal form transitions 
in the phase diagram \cite{KurodaLacmann}.

The above consideration regards only  the surface structures of faces. 
The situation becomes more interesting when we also consider  
the surface structures of the edges and  vertices. 
There exhibit transitions that differ 
from those of the basal and prism faces, as we shown in the following section. 

\section{Quasi-liquid on the edges and vertices of snow crystals}
\label{sec: 3}

First, we approximate the energetic advantage represented by
the surface melting parameter 
for the edges and the vertices of snow crystals, 
$\Delta\sigma_{(\rm edge\hspace{0.1cm} \&\hspace{0.1cm} vertex)}$. 
In the process of snow crystal formation, the edges and  vertices 
are formed as the intersections of metastable and unstable faces, 
such as $(11\bar{2}0)$, $(10\bar{1}1)$ and $(11{\bar 2}1)$.
For this reason, we approximate 
$\Delta\sigma_{(\rm edge\hspace{0.1cm}\&\hspace{0.1cm} vertex)}$ 
by taking the average of $\Delta\sigma$ for the intersecting faces.
It is known from the molecular structure of ice that 
these metastable and unstable faces
are rough and have large broken bond densities
in comparison with the faces $(0001)$ and $(10{\bar 1}0)$.
Thus, from Eq.(\ref{eq: Dsigma}), we see that  
$\Delta\sigma_{(\rm edge \hspace{0.1cm}\& \hspace{0.1cm}vertex)}$ 
is much larger than  
$\Delta \sigma_{(0001)}$ and $\Delta \sigma_{(10{\bar 1}0)}$:
\begin{eqnarray}
\Delta \sigma_{(0001)}< \Delta \sigma_{(10\bar{1}0)} 
< \Delta\sigma_{(\rm edge\hspace{0.1cm} \&\hspace{0.1cm} vertex)}.
\end{eqnarray}
This implies that for the edges and vertices, 
it is more energetically advantageous to have a surface  
transition from solid to liquid, and thus that they have a 
greater affinity for quasi-liquid. 
It is suggested by  Eqs.(\ref{eq: Deqflat1}) and (\ref{eq: Deqflat2}) 
that there exists quasi-liquid on the edges and  vertices   
at lower temperatures 
than on the faces $(0001)$ and $(10{\bar 1}0)$.

\begin{figure*}
\includegraphics[width=15cm]{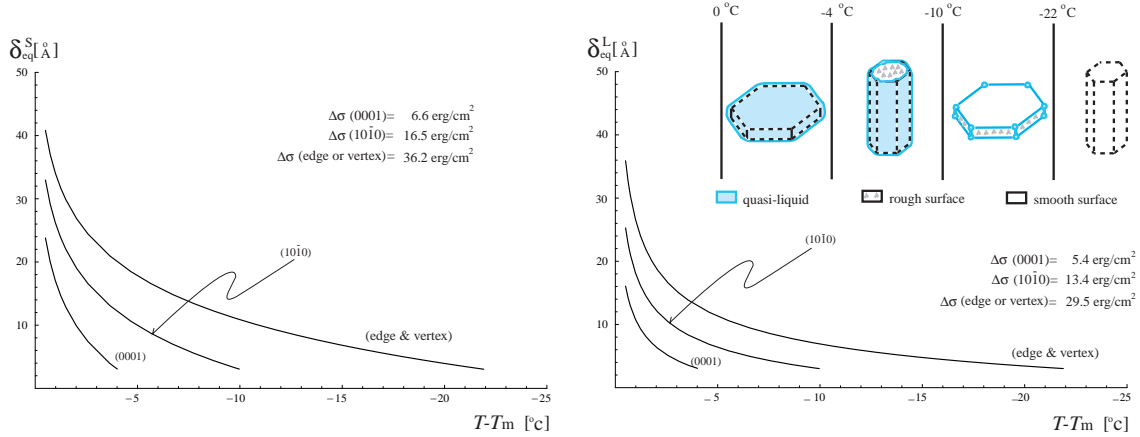}
\caption{\label{fig: Deq}
The equilibrium thickness of a quasi-liquid layer 
for $W^{\rm S}(\delta)$ and $W^{\rm L}(\delta)$ 
as functions of  temperature for various surface orientations. 
We chose $A=10 \stackrel{\circ}{\rm A}$, 
corresponding to the characteristic coherence length of 
a water molecule in the quasi-liquid \cite{Flet}.  
This is also the cluster size in water \cite{Nenow}.
The quantity $\Delta \sigma$ is computed using the   
monolayer thickness $(3 \stackrel{\circ}{\rm A})$
at $-4^{\circ}{\rm C}$ 
, $-10^{\circ}{\rm C}$ and $-22^{\circ}{\rm C}$.
( From Eq.(\ref{eq: Dsigma}), we obtain 
 $\Delta \sigma (0001) = 31 $ ${\rm erg/cm^2}$ , 
$\Delta \sigma (10{\bar 1}0)= 37 $ ${\rm erg/cm^2}$ 
and $\Delta \sigma (11{\bar 2}0) = 54 $ ${\rm erg/cm^2}$, 
where we use experimental values 
$\sigma_{\rm \ell v}= 76 {\rm erg/cm^2}$ at $0^{\circ}{\rm C}$,
$Q_{\rm s}= 8.48 \times 10^{-13} {\rm erg/molecule} $, 
and
$Q_{\rm m}= 1.0 \times 10^{-13} {\rm erg/molecule} $
 \cite{KurodaLacmann} . 
These values of $\Delta\sigma$ are not accurate,  
because $\sigma_{\rm vs}$ for an actual relaxed vapor-solid interface 
is less than $\rho Q_{\rm s}/(2N_{\rm B})$ 
for a freshly cut surface \cite{Cloud}.
Also, for actual values of $\sigma_{\rm s\ell}$, 
we must take into account  the fact 
that the water-ice interface has a diffuse structure throughout 
the thickness of several molecular layers \cite{Nada}.   
However, it is certain that 
$\Delta\sigma_{(0001)}<\Delta\sigma_{(10{\bar 1}0)}
<\Delta\sigma_{\rm (edge \hspace{1.5mm}\& \hspace{1.5mm}vertex)}$.)}   
\end{figure*}

In Fig.\ref{fig: Deq}, we present surface structures of snow crystals 
in which the quasi-liquid of the edges and  vertices 
are taken into account.
We find the following: 
(I) from $0^{\circ}$C to $-4^{\circ}$C, 
all the faces, edges and  vertices are covered with  quasi-liquid; 
(II) from $-4^{\circ}$C to $-10^{\circ}$C, 
the faces $(10\bar{1}0)$, edges and vertices 
are covered with  quasi-liquid, 
while the faces $(0001)$ have no quasi-liquid and are rough;
(III) from $-10^{\circ}$C to $-22^{\circ}$C, 
the edges and  vertices are covered with quasi-liquid, 
while the faces $(0001)$ are smooth and 
the faces $(10{\bar 1}0)$ are rough, both without quasi-liquid; 
(IV) below $-22^{\circ}$C, no quasi-liquid exists.
In the following sections, we examine the stability of the
quasi-liquids on the edges and  vertices in a vapor environment.

\section {Quasi-liquid layer on a curved solid surface}
\label{sec: 4}
To investigate the curvature effect on quasi-liquid 
at the edges and vertices,
we consider here three- and two-dimensional ice particles covered with  
quasi-liquid\footnote{Other curvature models of  surface melting  
are presented in Refs. \cite{BD} and \cite{NT}.}.
We assume that the particles are in a vapor environment 
and have radius $h+\delta$, 
where $h$ is the radius of the solid core and 
$\delta$ is the thickness of the quasi-liquid layer.  
Although the main reason we consider this system is to 
examine curvature effects on a quasi-liquid, 
the system itself has meaning as a model of ice nucleation 
in a vapor environment. 

We assume a free energy of the above described system as   
\begin{eqnarray}
&&\Delta G_{\rm particle}(h,\delta) 
=2^{D-1}\pi \Biggl[ h^{D-1}(\sigma_{\rm s \ell}
 +\Delta \sigma W(\delta))+(h+\delta)^{D-1} \sigma_{\ell {\rm v}} 
\nonumber\\ 
&&\hspace{20ex}
\left.
-\frac{1}{D}\left\{ (h+\delta)^D -h^D \right\}
  \frac{\mu_{\rm v}-\mu_{\ell}}{V_{\rm q}}
 -\frac{1}{D}h^D\frac{\mu_{\rm v}-\mu_{\rm s}}{V_{\rm s}}\right],
\label{eq: Gsphere}
\end{eqnarray}
where $D=3$ for three-dimensional particles and
$D=2$ for two-dimensional particles, 
$\mu_{\rm v}$, $\mu_{\rm \ell}$ and $\mu_{\rm s}$ are 
the chemical potentials per molecule for the bulk vapor, bulk liquid and  
bulk solid, and 
$V_{\rm q}$ and $V_{\rm s}$ 
are the volumes per molecule for the quasi-liquid and the solid.  
The first two terms on the right-hand side of Eq.(\ref{eq: Gsphere})
are the surface energies of the system, 
and the last two terms are the bulk contributions.  
The surface energy contribution must coincide with
$2^{D-1}\pi h^{D-1}\sigma_{\rm vs}$ for $\delta=0$ and 
$2^{D-1}\pi h^{D-1} \sigma_{\rm s\ell}
+2^{D-1}\pi(h+\delta)^{D-1}\sigma_{\ell v}$  
for $\delta \rightarrow \infty$. 
Thus $W(0)=1$ and $W(\infty)=0$. 
As in Eq.(\ref{eq: freeflat}), we use two types of $W(\delta)$: 
$W_{\rm S}(\delta)={\rm exp}(-\delta/A)$ and  
$W_{\rm L}(\delta)=(1+\delta/A)^{-2}$. 
The quantity $\Delta G_{\rm particle}(h,\delta)/(2^{D-1}\pi h^{D-1})$ 
reduces to  Eq.(\ref{eq: freeflat})
when $\mu_{\rm v}=\mu_{\rm s}$ and $h\rightarrow \infty$.
Since we are interested in the growth of ice in a vapor environment,
we assume $\mu_{\rm v}>\mu_{\rm s}$ 
vand $\mu_{\rm \ell}> \mu_{\rm s}$ in the following.

The equilibrium condition is given by the equations  
\begin{eqnarray}
\frac{\partial \Delta G_{\rm particle}}{\partial \delta} &=& 0,
\label{eq: CondeqD}
\\
\frac{\partial \Delta G_{\rm particle}}{\partial h}&=& 0.
\label{eq: CondeqH}
\end{eqnarray}
In equilibrium, we find a useful relation between $\delta$ and
$h$. 
From Eq.(\ref{eq: CondeqD}), we have
\begin{equation}
(D-1)(h+\delta)^{D-2}\sigma_{\rm \ell v}-(h+\delta)^{D-1}
\frac{\mu_{\rm v}-\mu_{\rm \ell}}{V_{\rm q}}
=-h^{D-1}\Delta\sigma\frac{dW(\delta)}{d\delta}.
\label{eq: CondD2}
\end{equation}
Thus, Eq.(\ref{eq: CondeqH}) becomes  
\begin{eqnarray}
(D-1)W(\delta)-h\frac{dW(\delta)}{d\delta}
=\frac{h}{\Delta \sigma}
\left(
\frac{\mu_{\rm v}-\mu_{\rm s}}{V_{\rm s}}-
\frac{\mu_{\rm v}-\mu_{\rm \ell}}{V_{\rm q}} -
\frac{(D-1)\sigma_{\rm s\ell}}{h}
\right).
\label{eq: CondH2}
\end{eqnarray}
This equation determines $\delta$ as a function of $h$.
If we use $W(\delta)=W_{\rm S}(\delta)$, we obtain 
\begin{equation}
 \delta^{\rm S}(h) = -A {\rm ln}\left[\frac {h}{(D-1)A +h}\right]
          -A{\rm ln}\left[\frac{A \lambda(h)}{\Delta \sigma}\right],
\label{eq: deltah1}
\end{equation}
and if we use $W(\delta)=W_{\rm L}(\delta)$, we obtain  
\begin{equation}
 \delta^{\rm L}(h) =-A + A\left( \chi (h,\lambda(h)) 
+\frac{(D-1)\Delta \sigma}{3h\lambda(h)}
\frac{1}{\chi(h,\lambda(h))}\right),
\label{eq: deltah2}
\end{equation}
where $\lambda(h)$ and $\chi(h,\lambda)$ are defined by 
\begin{eqnarray}
 \lambda(h)=\frac{\mu_{\ell}-\mu_{\rm s}}{V_{\rm q}} 
+\left( \frac{1}{V_{\rm s}} -\frac{1}{V_{\rm q}} \right)
(\mu_{\rm v} -\mu_{\rm s}) -\frac{(D-1) \sigma_{s\ell}}{h},
\label{eq: lambda}
\end{eqnarray}
and
\begin{eqnarray}
 \chi(h,\lambda)=\left\{ \frac{\Delta \sigma}{A \lambda}
\left(1+\sqrt{1-\frac{(D-1)^3 A^{2}\Delta\sigma}{27h^{3}\lambda}}\right)
\right\}^{\frac{1}{3}}.
\label{eq: chi}
\end{eqnarray}
The above functions $\delta^{\rm S}(h)$ and $\delta^{\rm L}(h)$ diverge at
$h=h_{\rm c}\equiv (D-1)\sigma_{\rm  s\ell}/
\left\{(\mu_{\rm v}-\mu_{\rm s})/V_{\rm s}
-(\mu_{\rm v}-\mu_{\ell})/V_{\rm q}\right\}$, and when $h>h_{\rm c}$,
they  monotonically decrease and approach the following 
$\delta^{\rm S}(\infty)$
and $\delta^{\rm L}(\infty)$ asymptotically:   
\begin{eqnarray}
\delta^{\rm S}(\infty) &=& 
-A{\rm ln}\left[ \frac{A\lambda(\infty)}{\Delta \sigma}\right],
\label{eq: hinf1} 
\\
\delta^{\rm L}(\infty) &=&
-A+\left( \frac{2A^2\Delta \sigma}
{\lambda(\infty)}\right)^{\frac{1}{3}}. 
\label{eq: hinf2}
\end{eqnarray}
No real solution of $\delta^{\rm S}(h)$ and
$\delta^{\rm L}(h)$ exists for $0<h<h_{\rm c}$.

When $V_{\rm s} > V_{\rm q}$, which holds for ice and water 
(and quasi-liquid), the following relations hold: 
\begin{eqnarray}
h_{\rm c} > (D-1)V_{\rm s}\sigma_{\rm s\ell}/(\mu_{\ell}-\mu_{\rm s}), 
\quad
\delta^{\rm S}(\infty)>\delta^{\rm S}_{\rm eq}, 
\quad 
\delta^{\rm L}(\infty)>\delta^{\rm L}_{\rm eq}. 
\end{eqnarray}
The quantities  $\delta^{\rm S}_{\rm eq}$ and $\delta^{\rm L}_{\rm eq}$ 
are given in 
Eqs.(\ref{eq: Deqflat1}) and (\ref{eq: Deqflat2}).  
Thus we find the following two properties of $h$ and $\delta$.
One is that $h > (D-1)V_{\rm s}\sigma_{\rm s\ell}/(\mu_{\ell}-\mu_{\rm s})$
in equilibrium.
For $D=3$, this means that the equilibrium radius of 
the solid part of the particle ,$h$, 
is always larger than the critical radius of ice nucleation in
supercooled water. 
The other is that 
$\delta^{\rm S}(h)$ in Eq.(\ref{eq: deltah1}) 
and $\delta^{\rm L}(h)$ in Eq.(\ref{eq: deltah2}) 
are always greater than 
$\delta_{\rm eq}^{\rm S}$ in Eq.(\ref{eq: Deqflat1}) 
and $\delta_{\rm eq}^{\rm L}$ in Eq.(\ref{eq: Deqflat2}), respectively.
This implies that the equilibrium thickness $\delta$ on a curved surface 
is greater than that on a plane surface, $\delta_{\rm eq}$ .

\begin{figure}
\includegraphics[width=8cm]{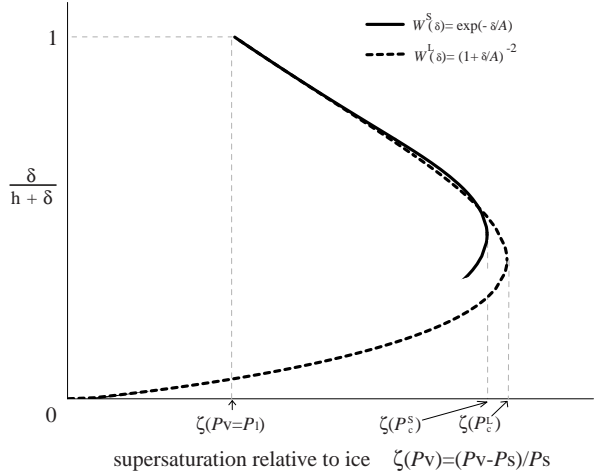}
\caption{\label{fig: sphereeq}
The equilibrium states of a quasi-liquid 
on a curved surface as a function of the supersaturation relative to ice.
For this figure, we set $ T = -15^{\circ}{\rm C}$, 
$\Delta \sigma= 36.2 $ ${\rm erg/cm^2}$, $A=10\stackrel{\circ}{\rm A}$ 
and $V_{\rm q}=V_{\rm s}=30 {\stackrel{\circ}{\rm A}}^{3}$.
From experimental values for water and ice, 
we have   
$\sigma_{\rm s\ell}=24.25   $ ${\rm erg/cm^2}$ and 
$\sigma_{\rm \ell v}= 78.25 $ ${\rm erg/cm^2}$
at $-15^{\circ}{\rm C}$ \cite{Cloud}.}
\end{figure}

We solved Eqs.(\ref{eq: CondeqD}) and (\ref{eq: CondeqH})  
numerically. 
In Fig.\ref{fig: sphereeq},  we plot the ratio 
$\delta/(h+\delta)$ in equilibrium 
as a function  
$(P_{\rm v}-P_{\rm s})/P_{\rm s}$, where $P_{\rm v}$ 
is the actual vapor pressure and 
$P_{\rm \ell}$ and $P_{\rm s}$ 
are the equilibrium vapor pressures 
for the bulk liquid and the bulk solid, respectively.
To obtain Fig.\ref{fig: sphereeq}, we have used the relations  
\begin{eqnarray}
\mu_{\rm v}(T,P_{\rm v})&=&\mu_{\rm v}(T)+kT{\rm ln}P_{\rm v}, 
\label{eq: mup1}
\\
\mu_{\rm \ell}(T,P_{\rm \ell})&=&\mu_{\rm v}(T,P_{\rm \ell})
=\mu_{\rm v}(T)+kT{\rm ln}P_{\rm \ell} ,
\label{eq: mup2}
\\
\mu_{\rm s}(T,P_{\rm s})&=&\mu_{\rm v}(T,P_{\rm s})
=\mu_{\rm v}(T)+kT{\rm ln}P_{\rm s}, 
\label{eq: mup3}
\end{eqnarray}
where $k$ is the Boltzmann constant and 
$\mu_{\rm v}(T)$ is the chemical potential of vapor at $P_{\rm v}=1$. 
When $P_{\ell}> P_{\rm v}>P_{\rm s}$, 
the equilibrium state is uniquely determined. 
When $P_{\rm v} \geq P_{\ell}$, in addition to this state,  
another equilibrium state 
that has a thicker quasi-liquid layer appears.
This state appears because above $P_{\rm \ell}$ 
it is possible not only for the solid but also for the liquid 
to grow in the vapor environment.  
As the supersaturation $P_{\rm v}$ increases, 
these two equilibrium states approach each other, and
at $P_{\rm v}=P_{\rm c}$ they coincide. 
Above $P_{\rm c}$ no equilibrium state exists.
All the equilibrium states obtained here are labile equilibrium states. 
From the viewpoint of ice nucleation in a vapor environment, 
they all correspond to a critical radius of nucleation.

\section{Instability of the quasi-liquid remaining on the edge of
 the crystal}    
\label{sec: 5}

\begin{figure}
\includegraphics[width=8cm]{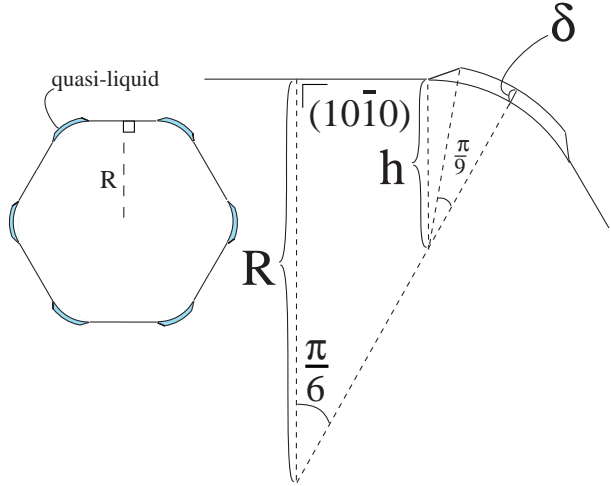}
\caption{\label{fig: prism}The quasi-liquid remaining on an edge.}
\end{figure}

Let us now consider a snow crystal whose edges 
are covered with quasi-liquid and 
whose faces have no  quasi-liquid layer (see Fig.\ref{fig: Deq}).
For the sake of simplicity, 
we consider only the six edges 
at which two different prism faces meet (see fig.\ref{fig: prism}).
For other edges and vertices, similar analyses can be carried out 
and yield qualitatively similar results.

The six prism faces $(10{\bar 1}0)$ grow slowly
in comparison with the edges, 
and therefore we use the approximation 
that their speed of growth is zero. 
As shown in Fig.\ref{fig: prism}, 
we also simplify the shape of the quasi-liquid layer on the edges. 
This simplification is justified when $\delta<h$.
For this system, the free energy of the snow crystal is given as 
\begin{eqnarray}
\Delta G_{\rm prism}(h,\delta)/12
&=&\left[
\frac{R}{\sqrt{3}}\sigma_{(10{\bar 1}0)}
-\frac{R^2}{2{\sqrt{3}}}\frac{\mu_{\rm v} -\mu_{\rm s}}{V_{\rm s}}\right]
-\left[\frac{h}{\sqrt{3}}\sigma_{(10{\bar 1}0)}
-\frac{h^2}{2{\sqrt{3}}}
\frac{\mu_{\rm v} -\mu_{\rm s}}{ V_{\rm s}}\right] \nonumber\\
\hspace{-0.8cm}
&+& 
\left[
\frac{\pi}{6}h(\sigma_{\rm s\ell(edge)}+\Delta\sigma_{\rm (edge)}
 W(\delta)) \right.
\nonumber\\
&&\hspace{-20ex}
+\left\{
\frac{\pi}{9}(h+\delta)+\sqrt{h^2+(h+\delta)^{2}-2h(h+\delta)
\cos \left(\frac{\pi}{18}\right)}
\right\} \sigma_{\rm \ell v}
\nonumber\\
&&\hspace{-20ex}
\left.
- \left\{\frac{\pi}{18}((h+\delta)^2-h^2)
+\frac{1}{2}\left( 
h(h+\delta)\sin\left(\frac{\pi}{18}\right)-\frac{\pi}{18} h^2\right)
\right\} 
 \frac{ \mu_{\rm v}- \mu_{\rm \ell}}{V_{\rm q}}
-\frac{\pi}{12}h^2\frac {\mu_{\rm v} -\mu_{\rm s}}{V_{\rm s}} 
\right], 
\label{eq: Gprism}
\end{eqnarray}
where 
$\sigma_{(10{\bar 1}0)}$ is the 
surface energy for a  prism face that has no quasi-liquid layer 
and adsorbs ${\rm H_{2}O}$ molecules, and 
$\sigma_{\rm s\ell(edge)}$ and $\Delta\sigma_{\rm(edge)}$ are 
the surface energy of the solid-liquid interface and 
the energetic advantage of surface melting 
(given by Eq.(\ref{eq: Dsigma})) for the surface of the edge, respectively.
The curvature effects considered in the previous section 
are taken into account 
by the terms inside the last set of square brackets 
on the right-hand side of Eq.(\ref{eq: Gprism}). 
When $h>\delta$ and $(\frac{\delta}{h})^3 \sim 0$,
Eq.(\ref{eq: Gprism}) becomes  
\begin{eqnarray}
\hspace{-7ex}
\Delta G_{\rm prism}(h,\delta)/12
&=&\left[
\frac{R}{\sqrt{3}}\sigma_{(10{\bar 1}0)}
-\frac{R^2}{2{\sqrt{3}}}\frac{\mu_{\rm v} -\mu_{\rm s}}{V_{\rm s}}\right]
-\left[\frac{h}{\sqrt{3}}\sigma_{(10{\bar 1}0)}
-\frac{h^2}{2{\sqrt{3}}}
\frac{\mu_{\rm v} -\mu_{\rm s}}{ V_{\rm s}}\right] \nonumber\\
&+&\left[
\frac{\pi}{6}h(\sigma_{\rm s\ell(edge)}+\Delta\sigma_{\rm (edge)}
 W(\delta)) \right.
\nonumber\\
&+&\frac{\pi}{9}\left\{
(h+\delta)
+\frac{1}{2}\left\{
1+\frac{\delta}{2h}
+\frac{1}{4}\left\{ \left( \frac{\pi}{36} \right)^{-2} -1 \right\}
\left(\frac{\delta}{h}\right)^{2}
\right\}h\right\}
 \sigma_{\rm \ell v}
\nonumber\\
&&
\hspace{18ex}
\left.
- \frac{\pi}{36}(2\delta^2+5h\delta) 
 \frac{ \mu_{\rm v}- \mu_{\rm \ell}}{V_{\rm q}}
-\frac{\pi}{12}h^2\frac {\mu_{\rm v} -\mu_{\rm s}}{V_{\rm s}} 
\right],
\label{eq: GprismA}
\end{eqnarray}
where
\begin{eqnarray}
\sqrt{h^2+(h+\delta)^2-2h(h+\delta)\cos(\frac{\pi}{18})}
=\sqrt{\delta^2+4h(h+\delta)\sin^2(\frac{\pi}{36})}
\nonumber\\
\simeq
2\sin(\frac{\pi}{36})h\left\{ 
1+\frac{\delta}{2h}
+\frac{1}{4}\left(\sin^{-2}(\frac{\pi}{36})-1\right)
\left(\frac{\delta}{h}\right)^2
\right\}
\nonumber,
\end{eqnarray} 
and we have used the approximations 
$\sin\left(\frac{\pi}{18}\right)\simeq \frac{\pi}{18}$ and   
$\sin\left(\frac{\pi}{36}\right)\simeq \frac{\pi}{36}$.  
The equilibrium condition is given by
\begin{eqnarray}
\frac{\partial \Delta G_{\rm prism}}{\partial \delta} = 0,
\label{eq: Eqprism1}
\\
\frac{\partial \Delta G_{\rm prism}}{\partial h} =  0.
\label{eq: Eqprism2}
\end{eqnarray}
From Eq.(\ref{eq: Eqprism2}), we find
\begin{eqnarray}
 h=h_{0}=\frac{
\frac{1}{\sqrt{3}}\sigma_{\rm (10{\bar 1}0)}
-\frac{\pi}{6}\sigma_{\rm vs(edge)}
}
{\left( \frac{1}{\sqrt{3}}-\frac{\pi}{6}\right)
\frac{\mu_{\rm v}-\mu_{\rm s}}{V_{\rm s}}
},
\label{eq: Hdelta}
\end{eqnarray}
where 
$\sigma_{\rm vs(edge)}$ is the surface energy of the vapor-solid
interface for the edge. 
In order for $h_{0}$ to be positive 
(which means that the curved edge is energetically stable) and for  
$\left(\frac{\partial \Delta G_{\rm prism}}{\partial \delta}\right)
^{h=h_{0}}_{\delta=0}<0$ to hold 
(which means that a curved edge with a quasi-liquid layer is 
energetically stable), 
$\sigma_{\rm (10{\bar 1}0)}$ must satisfy 
\begin{equation}
\sigma_{\rm (10{\bar 1}0)}>
  \frac{\sqrt{3}\pi}{6}\sigma_{\rm vs(edge)}
+ \frac{\frac{5}{6}\left( \frac{1}{\sqrt{3}}-\frac{\pi}{6} \right)
\sigma_{\rm \ell v}\frac{\mu_{\rm v}-\mu_{\rm s}}{V_{\rm s}}}
{-\Delta\sigma_{\rm (edge)}\frac{\partial W(0)}{\partial \delta}
+\frac{5}{6}\frac{\mu_{\rm v}-\mu_{\rm \ell}}{V_{\rm q}}}.
\label{eq: curvcon} 
\end{equation}
This equation gives a lower bound on  $\sigma_{\rm (10{\bar 1}0)}$.

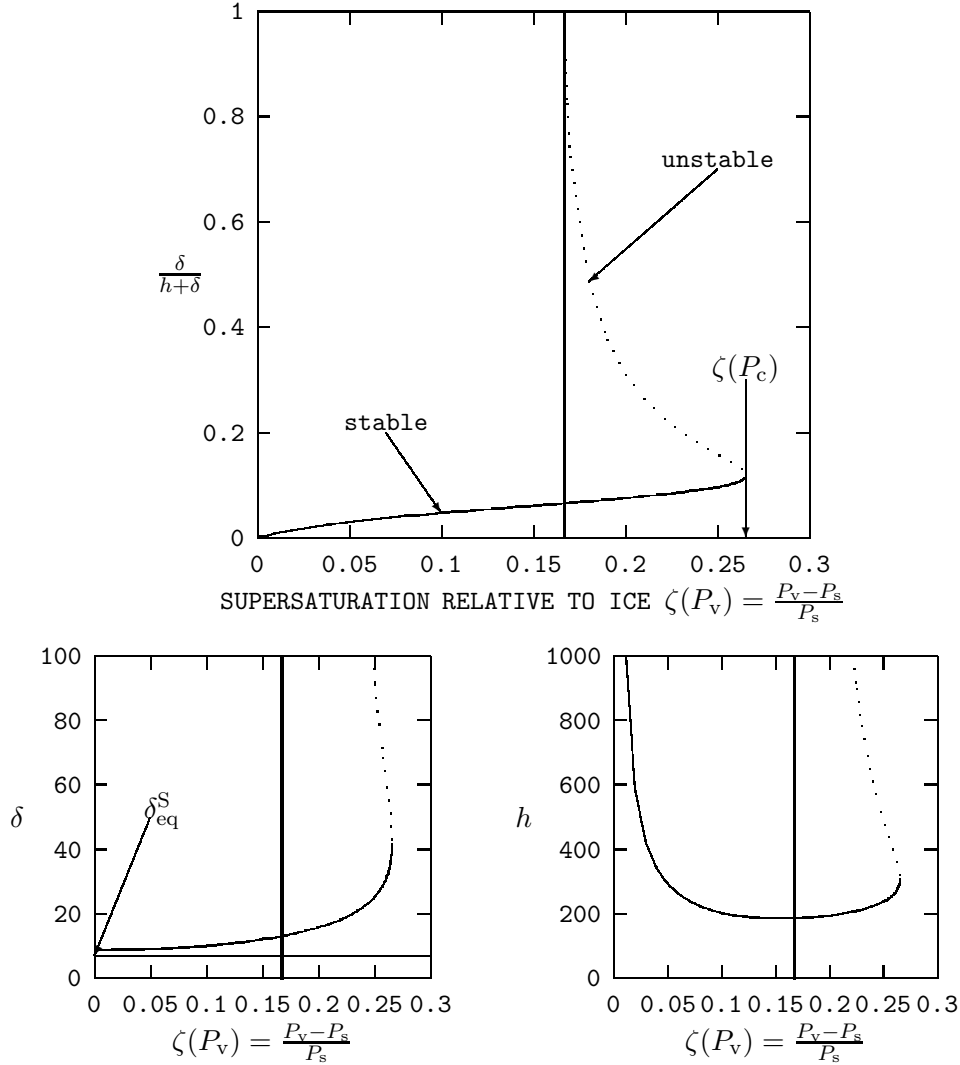
\begin{figure*}
\begin{center}
\setlength{\unitlength}{0.240900pt}
\ifx\plotpoint\undefined\newsavebox{\plotpoint}\fi
\sbox{\plotpoint}{\rule[-0.200pt]{0.400pt}{0.400pt}}%
\begin{picture}(1049,990)(0,0)
\font\gnuplot=cmtt10 at 10pt
\gnuplot
\sbox{\plotpoint}{\rule[-0.200pt]{0.400pt}{0.400pt}}%
\put(161.0,123.0){\rule[-0.200pt]{4.818pt}{0.400pt}}
\put(141,123){\makebox(0,0)[r]{0}}
\put(1008.0,123.0){\rule[-0.200pt]{4.818pt}{0.400pt}}
\put(161.0,288.0){\rule[-0.200pt]{4.818pt}{0.400pt}}
\put(141,288){\makebox(0,0)[r]{0.2}}
\put(1008.0,288.0){\rule[-0.200pt]{4.818pt}{0.400pt}}
\put(161.0,454.0){\rule[-0.200pt]{4.818pt}{0.400pt}}
\put(141,454){\makebox(0,0)[r]{0.4}}
\put(1008.0,454.0){\rule[-0.200pt]{4.818pt}{0.400pt}}
\put(161.0,619.0){\rule[-0.200pt]{4.818pt}{0.400pt}}
\put(141,619){\makebox(0,0)[r]{0.6}}
\put(1008.0,619.0){\rule[-0.200pt]{4.818pt}{0.400pt}}
\put(161.0,785.0){\rule[-0.200pt]{4.818pt}{0.400pt}}
\put(141,785){\makebox(0,0)[r]{0.8}}
\put(1008.0,785.0){\rule[-0.200pt]{4.818pt}{0.400pt}}
\put(161.0,950.0){\rule[-0.200pt]{4.818pt}{0.400pt}}
\put(141,950){\makebox(0,0)[r]{1}}
\put(1008.0,950.0){\rule[-0.200pt]{4.818pt}{0.400pt}}
\put(161.0,123.0){\rule[-0.200pt]{0.400pt}{4.818pt}}
\put(161,82){\makebox(0,0){0}}
\put(161.0,930.0){\rule[-0.200pt]{0.400pt}{4.818pt}}
\put(306.0,123.0){\rule[-0.200pt]{0.400pt}{4.818pt}}
\put(306,82){\makebox(0,0){0.05}}
\put(306.0,930.0){\rule[-0.200pt]{0.400pt}{4.818pt}}
\put(450.0,123.0){\rule[-0.200pt]{0.400pt}{4.818pt}}
\put(450,82){\makebox(0,0){0.1}}
\put(450.0,930.0){\rule[-0.200pt]{0.400pt}{4.818pt}}
\put(595.0,123.0){\rule[-0.200pt]{0.400pt}{4.818pt}}
\put(595,82){\makebox(0,0){0.15}}
\put(595.0,930.0){\rule[-0.200pt]{0.400pt}{4.818pt}}
\put(739.0,123.0){\rule[-0.200pt]{0.400pt}{4.818pt}}
\put(739,82){\makebox(0,0){0.2}}
\put(739.0,930.0){\rule[-0.200pt]{0.400pt}{4.818pt}}
\put(884.0,123.0){\rule[-0.200pt]{0.400pt}{4.818pt}}
\put(884,82){\makebox(0,0){0.25}}
\put(884.0,930.0){\rule[-0.200pt]{0.400pt}{4.818pt}}
\put(1028.0,123.0){\rule[-0.200pt]{0.400pt}{4.818pt}}
\put(1028,82){\makebox(0,0){0.3}}
\put(1028.0,930.0){\rule[-0.200pt]{0.400pt}{4.818pt}}
\put(161.0,123.0){\rule[-0.200pt]{208.860pt}{0.400pt}}
\put(1028.0,123.0){\rule[-0.200pt]{0.400pt}{199.224pt}}
\put(161.0,950.0){\rule[-0.200pt]{208.860pt}{0.400pt}}
\put(40,536){\makebox(0,0){$\frac{\delta}{h+\delta}$}}
\put(594,21){\makebox(0,0){SUPERSATURATION RELATIVE TO ICE $\zeta(P_{\rm v})=\frac{P_{\rm v}-P_{\rm s}}{P_{\rm s}}$}}
\put(363,305){\makebox(0,0){stable}}
\put(884,718){\makebox(0,0){unstable}}
\put(928,388){\makebox(0,0){$\zeta(P_{\rm c})$}}
\put(161.0,123.0){\rule[-0.200pt]{0.400pt}{199.224pt}}
\multiput(363.58,285.18)(0.499,-0.725){171}{\rule{0.120pt}{0.679pt}}
\multiput(362.17,286.59)(87.000,-124.590){2}{\rule{0.400pt}{0.340pt}}
\put(450,162){\vector(2,-3){0}}
\multiput(881.68,700.92)(-0.573,-0.500){351}{\rule{0.559pt}{0.120pt}}
\multiput(882.84,701.17)(-201.840,-177.000){2}{\rule{0.279pt}{0.400pt}}
\put(681,525){\vector(-4,-3){0}}
\put(928,371){\vector(0,-1){248}}
\put(161,123){\usebox{\plotpoint}}
\put(162,122.67){\rule{0.482pt}{0.400pt}}
\multiput(162.00,122.17)(1.000,1.000){2}{\rule{0.241pt}{0.400pt}}
\put(164,124.17){\rule{2.300pt}{0.400pt}}
\multiput(164.00,123.17)(6.226,2.000){2}{\rule{1.150pt}{0.400pt}}
\multiput(175.00,126.60)(2.090,0.468){5}{\rule{1.600pt}{0.113pt}}
\multiput(175.00,125.17)(11.679,4.000){2}{\rule{0.800pt}{0.400pt}}
\multiput(190.00,130.59)(3.159,0.477){7}{\rule{2.420pt}{0.115pt}}
\multiput(190.00,129.17)(23.977,5.000){2}{\rule{1.210pt}{0.400pt}}
\multiput(219.00,135.59)(3.159,0.477){7}{\rule{2.420pt}{0.115pt}}
\multiput(219.00,134.17)(23.977,5.000){2}{\rule{1.210pt}{0.400pt}}
\multiput(248.00,140.60)(4.137,0.468){5}{\rule{3.000pt}{0.113pt}}
\multiput(248.00,139.17)(22.773,4.000){2}{\rule{1.500pt}{0.400pt}}
\multiput(277.00,144.60)(4.137,0.468){5}{\rule{3.000pt}{0.113pt}}
\multiput(277.00,143.17)(22.773,4.000){2}{\rule{1.500pt}{0.400pt}}
\multiput(306.00,148.61)(6.044,0.447){3}{\rule{3.833pt}{0.108pt}}
\multiput(306.00,147.17)(20.044,3.000){2}{\rule{1.917pt}{0.400pt}}
\multiput(334.00,151.61)(6.267,0.447){3}{\rule{3.967pt}{0.108pt}}
\multiput(334.00,150.17)(20.767,3.000){2}{\rule{1.983pt}{0.400pt}}
\multiput(363.00,154.61)(6.267,0.447){3}{\rule{3.967pt}{0.108pt}}
\multiput(363.00,153.17)(20.767,3.000){2}{\rule{1.983pt}{0.400pt}}
\put(392,157.17){\rule{5.900pt}{0.400pt}}
\multiput(392.00,156.17)(16.754,2.000){2}{\rule{2.950pt}{0.400pt}}
\multiput(421.00,159.61)(6.267,0.447){3}{\rule{3.967pt}{0.108pt}}
\multiput(421.00,158.17)(20.767,3.000){2}{\rule{1.983pt}{0.400pt}}
\put(450,162.17){\rule{5.900pt}{0.400pt}}
\multiput(450.00,161.17)(16.754,2.000){2}{\rule{2.950pt}{0.400pt}}
\put(479,164.17){\rule{5.900pt}{0.400pt}}
\multiput(479.00,163.17)(16.754,2.000){2}{\rule{2.950pt}{0.400pt}}
\multiput(508.00,166.61)(6.267,0.447){3}{\rule{3.967pt}{0.108pt}}
\multiput(508.00,165.17)(20.767,3.000){2}{\rule{1.983pt}{0.400pt}}
\put(537,169.17){\rule{5.900pt}{0.400pt}}
\multiput(537.00,168.17)(16.754,2.000){2}{\rule{2.950pt}{0.400pt}}
\put(566,171.17){\rule{5.900pt}{0.400pt}}
\multiput(566.00,170.17)(16.754,2.000){2}{\rule{2.950pt}{0.400pt}}
\put(595,173.17){\rule{5.700pt}{0.400pt}}
\multiput(595.00,172.17)(16.169,2.000){2}{\rule{2.850pt}{0.400pt}}
\multiput(623.00,175.61)(6.267,0.447){3}{\rule{3.967pt}{0.108pt}}
\multiput(623.00,174.17)(20.767,3.000){2}{\rule{1.983pt}{0.400pt}}
\put(652,178.17){\rule{5.900pt}{0.400pt}}
\multiput(652.00,177.17)(16.754,2.000){2}{\rule{2.950pt}{0.400pt}}
\multiput(681.00,180.61)(6.267,0.447){3}{\rule{3.967pt}{0.108pt}}
\multiput(681.00,179.17)(20.767,3.000){2}{\rule{1.983pt}{0.400pt}}
\put(710,183.17){\rule{5.900pt}{0.400pt}}
\multiput(710.00,182.17)(16.754,2.000){2}{\rule{2.950pt}{0.400pt}}
\multiput(739.00,185.61)(6.267,0.447){3}{\rule{3.967pt}{0.108pt}}
\multiput(739.00,184.17)(20.767,3.000){2}{\rule{1.983pt}{0.400pt}}
\multiput(768.00,188.61)(6.267,0.447){3}{\rule{3.967pt}{0.108pt}}
\multiput(768.00,187.17)(20.767,3.000){2}{\rule{1.983pt}{0.400pt}}
\put(797,190.67){\rule{3.373pt}{0.400pt}}
\multiput(797.00,190.17)(7.000,1.000){2}{\rule{1.686pt}{0.400pt}}
\put(811,192.17){\rule{3.100pt}{0.400pt}}
\multiput(811.00,191.17)(8.566,2.000){2}{\rule{1.550pt}{0.400pt}}
\put(826,194.17){\rule{2.900pt}{0.400pt}}
\multiput(826.00,193.17)(7.981,2.000){2}{\rule{1.450pt}{0.400pt}}
\put(840,196.17){\rule{3.100pt}{0.400pt}}
\multiput(840.00,195.17)(8.566,2.000){2}{\rule{1.550pt}{0.400pt}}
\put(855,198.17){\rule{2.900pt}{0.400pt}}
\multiput(855.00,197.17)(7.981,2.000){2}{\rule{1.450pt}{0.400pt}}
\put(869,200.17){\rule{3.100pt}{0.400pt}}
\multiput(869.00,199.17)(8.566,2.000){2}{\rule{1.550pt}{0.400pt}}
\put(884,202.17){\rule{1.500pt}{0.400pt}}
\multiput(884.00,201.17)(3.887,2.000){2}{\rule{0.750pt}{0.400pt}}
\put(891,203.67){\rule{1.686pt}{0.400pt}}
\multiput(891.00,203.17)(3.500,1.000){2}{\rule{0.843pt}{0.400pt}}
\put(898,205.17){\rule{1.500pt}{0.400pt}}
\multiput(898.00,204.17)(3.887,2.000){2}{\rule{0.750pt}{0.400pt}}
\put(905,207.17){\rule{1.500pt}{0.400pt}}
\multiput(905.00,206.17)(3.887,2.000){2}{\rule{0.750pt}{0.400pt}}
\put(912,208.67){\rule{0.723pt}{0.400pt}}
\multiput(912.00,208.17)(1.500,1.000){2}{\rule{0.361pt}{0.400pt}}
\put(915,209.67){\rule{0.723pt}{0.400pt}}
\multiput(915.00,209.17)(1.500,1.000){2}{\rule{0.361pt}{0.400pt}}
\put(918,211.17){\rule{0.700pt}{0.400pt}}
\multiput(918.00,210.17)(1.547,2.000){2}{\rule{0.350pt}{0.400pt}}
\put(921,213.17){\rule{0.700pt}{0.400pt}}
\multiput(921.00,212.17)(1.547,2.000){2}{\rule{0.350pt}{0.400pt}}
\put(924,215.17){\rule{0.700pt}{0.400pt}}
\multiput(924.00,214.17)(1.547,2.000){2}{\rule{0.350pt}{0.400pt}}
\put(161.0,123.0){\usebox{\plotpoint}}
\put(927,217.67){\rule{0.241pt}{0.400pt}}
\multiput(927.00,217.17)(0.500,1.000){2}{\rule{0.120pt}{0.400pt}}
\put(927.0,217.0){\usebox{\plotpoint}}
\put(928,219){\usebox{\plotpoint}}
\put(928,219){\usebox{\plotpoint}}
\put(928.0,219.0){\rule[-0.200pt]{0.400pt}{0.482pt}}
\put(644,874){\usebox{\plotpoint}}
\multiput(644,874)(0.902,-20.736){4}{\usebox{\plotpoint}}
\multiput(647,805)(0.882,-20.737){2}{\usebox{\plotpoint}}
\multiput(649,758)(1.677,-20.688){2}{\usebox{\plotpoint}}
\multiput(652,721)(3.038,-20.532){9}{\usebox{\plotpoint}}
\multiput(681,525)(6.302,-19.776){5}{\usebox{\plotpoint}}
\multiput(710,434)(9.681,-18.360){3}{\usebox{\plotpoint}}
\multiput(739,379)(12.592,-16.500){2}{\usebox{\plotpoint}}
\multiput(768,341)(14.676,-14.676){2}{\usebox{\plotpoint}}
\put(800.07,309.37){\usebox{\plotpoint}}
\put(816.13,296.24){\usebox{\plotpoint}}
\put(832.93,284.05){\usebox{\plotpoint}}
\put(850.35,272.79){\usebox{\plotpoint}}
\put(867.90,261.71){\usebox{\plotpoint}}
\put(886.12,251.79){\usebox{\plotpoint}}
\put(904.14,241.49){\usebox{\plotpoint}}
\put(921.63,230.37){\usebox{\plotpoint}}
\put(928,221){\usebox{\plotpoint}}
\sbox{\plotpoint}{\rule[-0.400pt]{0.800pt}{0.800pt}}%
\put(642,123){\usebox{\plotpoint}}
\put(642.0,123.0){\rule[-0.400pt]{0.800pt}{199.224pt}}
\sbox{\plotpoint}{\rule[-0.500pt]{1.000pt}{1.000pt}}%
\end{picture} \\
 \vspace{0.5cm}
\setlength{\unitlength}{0.240900pt}
\ifx\plotpoint\undefined\newsavebox{\plotpoint}\fi
\sbox{\plotpoint}{\rule[-0.200pt]{0.400pt}{0.400pt}}%
\begin{picture}(750,629)(0,0)
\font\gnuplot=cmtt10 at 10pt
\gnuplot
\sbox{\plotpoint}{\rule[-0.200pt]{0.400pt}{0.400pt}}%
\put(161.0,123.0){\rule[-0.200pt]{4.818pt}{0.400pt}}
\put(141,123){\makebox(0,0)[r]{0}}
\put(669.0,123.0){\rule[-0.200pt]{4.818pt}{0.400pt}}
\put(161.0,224.0){\rule[-0.200pt]{4.818pt}{0.400pt}}
\put(141,224){\makebox(0,0)[r]{20}}
\put(669.0,224.0){\rule[-0.200pt]{4.818pt}{0.400pt}}
\put(161.0,325.0){\rule[-0.200pt]{4.818pt}{0.400pt}}
\put(141,325){\makebox(0,0)[r]{40}}
\put(669.0,325.0){\rule[-0.200pt]{4.818pt}{0.400pt}}
\put(161.0,427.0){\rule[-0.200pt]{4.818pt}{0.400pt}}
\put(141,427){\makebox(0,0)[r]{60}}
\put(669.0,427.0){\rule[-0.200pt]{4.818pt}{0.400pt}}
\put(161.0,528.0){\rule[-0.200pt]{4.818pt}{0.400pt}}
\put(141,528){\makebox(0,0)[r]{80}}
\put(669.0,528.0){\rule[-0.200pt]{4.818pt}{0.400pt}}
\put(161.0,629.0){\rule[-0.200pt]{4.818pt}{0.400pt}}
\put(141,629){\makebox(0,0)[r]{100}}
\put(669.0,629.0){\rule[-0.200pt]{4.818pt}{0.400pt}}
\put(161.0,123.0){\rule[-0.200pt]{0.400pt}{4.818pt}}
\put(161,82){\makebox(0,0){0}}
\put(161.0,609.0){\rule[-0.200pt]{0.400pt}{4.818pt}}
\put(249.0,123.0){\rule[-0.200pt]{0.400pt}{4.818pt}}
\put(249,82){\makebox(0,0){0.05}}
\put(249.0,609.0){\rule[-0.200pt]{0.400pt}{4.818pt}}
\put(337.0,123.0){\rule[-0.200pt]{0.400pt}{4.818pt}}
\put(337,82){\makebox(0,0){0.1}}
\put(337.0,609.0){\rule[-0.200pt]{0.400pt}{4.818pt}}
\put(425.0,123.0){\rule[-0.200pt]{0.400pt}{4.818pt}}
\put(425,82){\makebox(0,0){0.15}}
\put(425.0,609.0){\rule[-0.200pt]{0.400pt}{4.818pt}}
\put(513.0,123.0){\rule[-0.200pt]{0.400pt}{4.818pt}}
\put(513,82){\makebox(0,0){0.2}}
\put(513.0,609.0){\rule[-0.200pt]{0.400pt}{4.818pt}}
\put(601.0,123.0){\rule[-0.200pt]{0.400pt}{4.818pt}}
\put(601,82){\makebox(0,0){0.25}}
\put(601.0,609.0){\rule[-0.200pt]{0.400pt}{4.818pt}}
\put(689.0,123.0){\rule[-0.200pt]{0.400pt}{4.818pt}}
\put(689,82){\makebox(0,0){0.3}}
\put(689.0,609.0){\rule[-0.200pt]{0.400pt}{4.818pt}}
\put(161.0,123.0){\rule[-0.200pt]{127.195pt}{0.400pt}}
\put(689.0,123.0){\rule[-0.200pt]{0.400pt}{121.895pt}}
\put(161.0,629.0){\rule[-0.200pt]{127.195pt}{0.400pt}}
\put(40,376){\makebox(0,0){$\delta$}}
\put(425,21){\makebox(0,0){$\zeta(P_{\rm v})=\frac{P_{\rm v}-P_{\rm s}}{P_{\rm s}}$}}
\put(267,386){\makebox(0,0){$\delta_{\rm eq}^{\rm S} $}}
\put(161.0,123.0){\rule[-0.200pt]{0.400pt}{121.895pt}}
\multiput(247.92,371.47)(-0.499,-1.241){173}{\rule{0.120pt}{1.091pt}}
\multiput(248.17,373.74)(-88.000,-215.736){2}{\rule{0.400pt}{0.545pt}}
\put(161,158){\vector(-1,-2){0}}
\put(161,167){\usebox{\plotpoint}}
\put(179,166.67){\rule{4.095pt}{0.400pt}}
\multiput(179.00,166.17)(8.500,1.000){2}{\rule{2.048pt}{0.400pt}}
\put(161.0,167.0){\rule[-0.200pt]{4.336pt}{0.400pt}}
\put(231,167.67){\rule{4.336pt}{0.400pt}}
\multiput(231.00,167.17)(9.000,1.000){2}{\rule{2.168pt}{0.400pt}}
\put(196.0,168.0){\rule[-0.200pt]{8.431pt}{0.400pt}}
\put(267,168.67){\rule{4.095pt}{0.400pt}}
\multiput(267.00,168.17)(8.500,1.000){2}{\rule{2.048pt}{0.400pt}}
\put(284,169.67){\rule{4.336pt}{0.400pt}}
\multiput(284.00,169.17)(9.000,1.000){2}{\rule{2.168pt}{0.400pt}}
\put(302,170.67){\rule{4.095pt}{0.400pt}}
\multiput(302.00,170.17)(8.500,1.000){2}{\rule{2.048pt}{0.400pt}}
\put(319,171.67){\rule{4.336pt}{0.400pt}}
\multiput(319.00,171.17)(9.000,1.000){2}{\rule{2.168pt}{0.400pt}}
\put(337,173.17){\rule{3.700pt}{0.400pt}}
\multiput(337.00,172.17)(10.320,2.000){2}{\rule{1.850pt}{0.400pt}}
\put(355,175.17){\rule{3.500pt}{0.400pt}}
\multiput(355.00,174.17)(9.736,2.000){2}{\rule{1.750pt}{0.400pt}}
\put(372,177.17){\rule{3.700pt}{0.400pt}}
\multiput(372.00,176.17)(10.320,2.000){2}{\rule{1.850pt}{0.400pt}}
\put(390,179.17){\rule{3.500pt}{0.400pt}}
\multiput(390.00,178.17)(9.736,2.000){2}{\rule{1.750pt}{0.400pt}}
\multiput(407.00,181.61)(3.811,0.447){3}{\rule{2.500pt}{0.108pt}}
\multiput(407.00,180.17)(12.811,3.000){2}{\rule{1.250pt}{0.400pt}}
\put(425,184.17){\rule{3.700pt}{0.400pt}}
\multiput(425.00,183.17)(10.320,2.000){2}{\rule{1.850pt}{0.400pt}}
\multiput(443.00,186.60)(2.382,0.468){5}{\rule{1.800pt}{0.113pt}}
\multiput(443.00,185.17)(13.264,4.000){2}{\rule{0.900pt}{0.400pt}}
\multiput(460.00,190.60)(2.528,0.468){5}{\rule{1.900pt}{0.113pt}}
\multiput(460.00,189.17)(14.056,4.000){2}{\rule{0.950pt}{0.400pt}}
\multiput(478.00,194.60)(2.382,0.468){5}{\rule{1.800pt}{0.113pt}}
\multiput(478.00,193.17)(13.264,4.000){2}{\rule{0.900pt}{0.400pt}}
\multiput(495.00,198.59)(1.935,0.477){7}{\rule{1.540pt}{0.115pt}}
\multiput(495.00,197.17)(14.804,5.000){2}{\rule{0.770pt}{0.400pt}}
\multiput(513.00,203.59)(1.935,0.477){7}{\rule{1.540pt}{0.115pt}}
\multiput(513.00,202.17)(14.804,5.000){2}{\rule{0.770pt}{0.400pt}}
\multiput(531.00,208.59)(1.255,0.485){11}{\rule{1.071pt}{0.117pt}}
\multiput(531.00,207.17)(14.776,7.000){2}{\rule{0.536pt}{0.400pt}}
\multiput(548.00,215.60)(1.212,0.468){5}{\rule{1.000pt}{0.113pt}}
\multiput(548.00,214.17)(6.924,4.000){2}{\rule{0.500pt}{0.400pt}}
\multiput(557.00,219.59)(0.933,0.477){7}{\rule{0.820pt}{0.115pt}}
\multiput(557.00,218.17)(7.298,5.000){2}{\rule{0.410pt}{0.400pt}}
\multiput(566.00,224.59)(0.933,0.477){7}{\rule{0.820pt}{0.115pt}}
\multiput(566.00,223.17)(7.298,5.000){2}{\rule{0.410pt}{0.400pt}}
\multiput(575.00,229.59)(0.671,0.482){9}{\rule{0.633pt}{0.116pt}}
\multiput(575.00,228.17)(6.685,6.000){2}{\rule{0.317pt}{0.400pt}}
\multiput(583.00,235.59)(0.762,0.482){9}{\rule{0.700pt}{0.116pt}}
\multiput(583.00,234.17)(7.547,6.000){2}{\rule{0.350pt}{0.400pt}}
\multiput(592.00,241.59)(0.495,0.489){15}{\rule{0.500pt}{0.118pt}}
\multiput(592.00,240.17)(7.962,9.000){2}{\rule{0.250pt}{0.400pt}}
\multiput(601.60,250.00)(0.468,0.627){5}{\rule{0.113pt}{0.600pt}}
\multiput(600.17,250.00)(4.000,3.755){2}{\rule{0.400pt}{0.300pt}}
\multiput(605.59,255.00)(0.477,0.599){7}{\rule{0.115pt}{0.580pt}}
\multiput(604.17,255.00)(5.000,4.796){2}{\rule{0.400pt}{0.290pt}}
\multiput(610.60,261.00)(0.468,0.920){5}{\rule{0.113pt}{0.800pt}}
\multiput(609.17,261.00)(4.000,5.340){2}{\rule{0.400pt}{0.400pt}}
\multiput(614.59,268.00)(0.477,0.933){7}{\rule{0.115pt}{0.820pt}}
\multiput(613.17,268.00)(5.000,7.298){2}{\rule{0.400pt}{0.410pt}}
\put(618.67,277){\rule{0.400pt}{0.964pt}}
\multiput(618.17,277.00)(1.000,2.000){2}{\rule{0.400pt}{0.482pt}}
\put(620.17,281){\rule{0.400pt}{1.100pt}}
\multiput(619.17,281.00)(2.000,2.717){2}{\rule{0.400pt}{0.550pt}}
\put(622.17,286){\rule{0.400pt}{1.500pt}}
\multiput(621.17,286.00)(2.000,3.887){2}{\rule{0.400pt}{0.750pt}}
\put(624.17,293){\rule{0.400pt}{1.700pt}}
\multiput(623.17,293.00)(2.000,4.472){2}{\rule{0.400pt}{0.850pt}}
\put(625.67,301){\rule{0.400pt}{3.373pt}}
\multiput(625.17,301.00)(1.000,7.000){2}{\rule{0.400pt}{1.686pt}}
\put(626.67,315){\rule{0.400pt}{0.482pt}}
\multiput(626.17,315.00)(1.000,1.000){2}{\rule{0.400pt}{0.241pt}}
\put(249.0,169.0){\rule[-0.200pt]{4.336pt}{0.400pt}}
\put(628.0,317.0){\rule[-0.200pt]{0.400pt}{3.854pt}}
\multiput(597,629)(1.798,-20.677){3}{\usebox{\plotpoint}}
\multiput(601,583)(2.015,-20.657){2}{\usebox{\plotpoint}}
\multiput(605,542)(2.780,-20.569){2}{\usebox{\plotpoint}}
\put(612.27,484.54){\usebox{\plotpoint}}
\multiput(614,469)(3.020,-20.535){2}{\usebox{\plotpoint}}
\put(619.87,422.77){\usebox{\plotpoint}}
\put(622.64,402.20){\usebox{\plotpoint}}
\put(625.22,381.61){\usebox{\plotpoint}}
\put(626.64,360.91){\usebox{\plotpoint}}
\put(628.00,340.33){\usebox{\plotpoint}}
\put(628,333){\usebox{\plotpoint}}
\sbox{\plotpoint}{\rule[-0.400pt]{0.800pt}{0.800pt}}%
\put(454,123){\usebox{\plotpoint}}
\put(454.0,123.0){\rule[-0.400pt]{0.800pt}{121.895pt}}
\sbox{\plotpoint}{\rule[-0.200pt]{0.400pt}{0.400pt}}%
\put(161,158){\usebox{\plotpoint}}
\put(161.0,158.0){\rule[-0.200pt]{127.195pt}{0.400pt}}
\end{picture}
\setlength{\unitlength}{0.240900pt}
\ifx\plotpoint\undefined\newsavebox{\plotpoint}\fi
\sbox{\plotpoint}{\rule[-0.200pt]{0.400pt}{0.400pt}}%
\begin{picture}(750,629)(0,0)
\font\gnuplot=cmtt10 at 10pt
\gnuplot
\sbox{\plotpoint}{\rule[-0.200pt]{0.400pt}{0.400pt}}%
\put(181.0,123.0){\rule[-0.200pt]{4.818pt}{0.400pt}}
\put(161,123){\makebox(0,0)[r]{0}}
\put(669.0,123.0){\rule[-0.200pt]{4.818pt}{0.400pt}}
\put(181.0,224.0){\rule[-0.200pt]{4.818pt}{0.400pt}}
\put(161,224){\makebox(0,0)[r]{200}}
\put(669.0,224.0){\rule[-0.200pt]{4.818pt}{0.400pt}}
\put(181.0,325.0){\rule[-0.200pt]{4.818pt}{0.400pt}}
\put(161,325){\makebox(0,0)[r]{400}}
\put(669.0,325.0){\rule[-0.200pt]{4.818pt}{0.400pt}}
\put(181.0,427.0){\rule[-0.200pt]{4.818pt}{0.400pt}}
\put(161,427){\makebox(0,0)[r]{600}}
\put(669.0,427.0){\rule[-0.200pt]{4.818pt}{0.400pt}}
\put(181.0,528.0){\rule[-0.200pt]{4.818pt}{0.400pt}}
\put(161,528){\makebox(0,0)[r]{800}}
\put(669.0,528.0){\rule[-0.200pt]{4.818pt}{0.400pt}}
\put(181.0,629.0){\rule[-0.200pt]{4.818pt}{0.400pt}}
\put(161,629){\makebox(0,0)[r]{1000}}
\put(669.0,629.0){\rule[-0.200pt]{4.818pt}{0.400pt}}
\put(181.0,123.0){\rule[-0.200pt]{0.400pt}{4.818pt}}
\put(181,82){\makebox(0,0){0}}
\put(181.0,609.0){\rule[-0.200pt]{0.400pt}{4.818pt}}
\put(266.0,123.0){\rule[-0.200pt]{0.400pt}{4.818pt}}
\put(266,82){\makebox(0,0){0.05}}
\put(266.0,609.0){\rule[-0.200pt]{0.400pt}{4.818pt}}
\put(350.0,123.0){\rule[-0.200pt]{0.400pt}{4.818pt}}
\put(350,82){\makebox(0,0){0.1}}
\put(350.0,609.0){\rule[-0.200pt]{0.400pt}{4.818pt}}
\put(435.0,123.0){\rule[-0.200pt]{0.400pt}{4.818pt}}
\put(435,82){\makebox(0,0){0.15}}
\put(435.0,609.0){\rule[-0.200pt]{0.400pt}{4.818pt}}
\put(520.0,123.0){\rule[-0.200pt]{0.400pt}{4.818pt}}
\put(520,82){\makebox(0,0){0.2}}
\put(520.0,609.0){\rule[-0.200pt]{0.400pt}{4.818pt}}
\put(604.0,123.0){\rule[-0.200pt]{0.400pt}{4.818pt}}
\put(604,82){\makebox(0,0){0.25}}
\put(604.0,609.0){\rule[-0.200pt]{0.400pt}{4.818pt}}
\put(689.0,123.0){\rule[-0.200pt]{0.400pt}{4.818pt}}
\put(689,82){\makebox(0,0){0.3}}
\put(689.0,609.0){\rule[-0.200pt]{0.400pt}{4.818pt}}
\put(181.0,123.0){\rule[-0.200pt]{122.377pt}{0.400pt}}
\put(689.0,123.0){\rule[-0.200pt]{0.400pt}{121.895pt}}
\put(181.0,629.0){\rule[-0.200pt]{122.377pt}{0.400pt}}
\put(40,376){\makebox(0,0){$h$}}
\put(435,21){\makebox(0,0){$\zeta(P_{\rm v})=\frac{P_{\rm v}-P_{\rm s}}{P_{\rm s}}$}}
\put(181.0,123.0){\rule[-0.200pt]{0.400pt}{121.895pt}}
\multiput(201.58,603.68)(0.494,-7.695){25}{\rule{0.119pt}{6.100pt}}
\multiput(200.17,616.34)(14.000,-197.339){2}{\rule{0.400pt}{3.050pt}}
\multiput(215.58,410.48)(0.495,-2.483){31}{\rule{0.119pt}{2.053pt}}
\multiput(214.17,414.74)(17.000,-78.739){2}{\rule{0.400pt}{1.026pt}}
\multiput(232.58,331.58)(0.495,-1.219){31}{\rule{0.119pt}{1.065pt}}
\multiput(231.17,333.79)(17.000,-38.790){2}{\rule{0.400pt}{0.532pt}}
\multiput(249.58,292.24)(0.495,-0.708){31}{\rule{0.119pt}{0.665pt}}
\multiput(248.17,293.62)(17.000,-22.620){2}{\rule{0.400pt}{0.332pt}}
\multiput(266.00,269.92)(0.529,-0.494){29}{\rule{0.525pt}{0.119pt}}
\multiput(266.00,270.17)(15.910,-16.000){2}{\rule{0.263pt}{0.400pt}}
\multiput(283.00,253.92)(0.779,-0.492){19}{\rule{0.718pt}{0.118pt}}
\multiput(283.00,254.17)(15.509,-11.000){2}{\rule{0.359pt}{0.400pt}}
\multiput(300.00,242.93)(1.022,-0.488){13}{\rule{0.900pt}{0.117pt}}
\multiput(300.00,243.17)(14.132,-8.000){2}{\rule{0.450pt}{0.400pt}}
\multiput(316.00,234.93)(1.485,-0.482){9}{\rule{1.233pt}{0.116pt}}
\multiput(316.00,235.17)(14.440,-6.000){2}{\rule{0.617pt}{0.400pt}}
\multiput(333.00,228.93)(1.823,-0.477){7}{\rule{1.460pt}{0.115pt}}
\multiput(333.00,229.17)(13.970,-5.000){2}{\rule{0.730pt}{0.400pt}}
\multiput(350.00,223.95)(3.588,-0.447){3}{\rule{2.367pt}{0.108pt}}
\multiput(350.00,224.17)(12.088,-3.000){2}{\rule{1.183pt}{0.400pt}}
\put(367,220.17){\rule{3.500pt}{0.400pt}}
\multiput(367.00,221.17)(9.736,-2.000){2}{\rule{1.750pt}{0.400pt}}
\put(384,218.17){\rule{3.500pt}{0.400pt}}
\multiput(384.00,219.17)(9.736,-2.000){2}{\rule{1.750pt}{0.400pt}}
\put(401,216.67){\rule{4.095pt}{0.400pt}}
\multiput(401.00,217.17)(8.500,-1.000){2}{\rule{2.048pt}{0.400pt}}
\put(469,216.67){\rule{4.095pt}{0.400pt}}
\multiput(469.00,216.17)(8.500,1.000){2}{\rule{2.048pt}{0.400pt}}
\put(486,217.67){\rule{4.095pt}{0.400pt}}
\multiput(486.00,217.17)(8.500,1.000){2}{\rule{2.048pt}{0.400pt}}
\put(503,219.17){\rule{3.500pt}{0.400pt}}
\multiput(503.00,218.17)(9.736,2.000){2}{\rule{1.750pt}{0.400pt}}
\multiput(520.00,221.61)(3.588,0.447){3}{\rule{2.367pt}{0.108pt}}
\multiput(520.00,220.17)(12.088,3.000){2}{\rule{1.183pt}{0.400pt}}
\multiput(537.00,224.61)(3.588,0.447){3}{\rule{2.367pt}{0.108pt}}
\multiput(537.00,223.17)(12.088,3.000){2}{\rule{1.183pt}{0.400pt}}
\put(554,226.67){\rule{1.927pt}{0.400pt}}
\multiput(554.00,226.17)(4.000,1.000){2}{\rule{0.964pt}{0.400pt}}
\put(562,228.17){\rule{1.700pt}{0.400pt}}
\multiput(562.00,227.17)(4.472,2.000){2}{\rule{0.850pt}{0.400pt}}
\multiput(570.00,230.61)(1.802,0.447){3}{\rule{1.300pt}{0.108pt}}
\multiput(570.00,229.17)(6.302,3.000){2}{\rule{0.650pt}{0.400pt}}
\multiput(579.00,233.61)(1.579,0.447){3}{\rule{1.167pt}{0.108pt}}
\multiput(579.00,232.17)(5.579,3.000){2}{\rule{0.583pt}{0.400pt}}
\multiput(587.00,236.61)(1.802,0.447){3}{\rule{1.300pt}{0.108pt}}
\multiput(587.00,235.17)(6.302,3.000){2}{\rule{0.650pt}{0.400pt}}
\multiput(596.00,239.60)(1.066,0.468){5}{\rule{0.900pt}{0.113pt}}
\multiput(596.00,238.17)(6.132,4.000){2}{\rule{0.450pt}{0.400pt}}
\put(604,243.17){\rule{1.100pt}{0.400pt}}
\multiput(604.00,242.17)(2.717,2.000){2}{\rule{0.550pt}{0.400pt}}
\multiput(609.00,245.61)(0.685,0.447){3}{\rule{0.633pt}{0.108pt}}
\multiput(609.00,244.17)(2.685,3.000){2}{\rule{0.317pt}{0.400pt}}
\multiput(613.00,248.61)(0.685,0.447){3}{\rule{0.633pt}{0.108pt}}
\multiput(613.00,247.17)(2.685,3.000){2}{\rule{0.317pt}{0.400pt}}
\multiput(617.00,251.60)(0.481,0.468){5}{\rule{0.500pt}{0.113pt}}
\multiput(617.00,250.17)(2.962,4.000){2}{\rule{0.250pt}{0.400pt}}
\put(621,255.17){\rule{0.482pt}{0.400pt}}
\multiput(621.00,254.17)(1.000,2.000){2}{\rule{0.241pt}{0.400pt}}
\put(623,257.17){\rule{0.482pt}{0.400pt}}
\multiput(623.00,256.17)(1.000,2.000){2}{\rule{0.241pt}{0.400pt}}
\put(624.67,259){\rule{0.400pt}{0.723pt}}
\multiput(624.17,259.00)(1.000,1.500){2}{\rule{0.400pt}{0.361pt}}
\put(626.17,262){\rule{0.400pt}{0.900pt}}
\multiput(625.17,262.00)(2.000,2.132){2}{\rule{0.400pt}{0.450pt}}
\put(628.17,266){\rule{0.400pt}{1.300pt}}
\multiput(627.17,266.00)(2.000,3.302){2}{\rule{0.400pt}{0.650pt}}
\put(418.0,217.0){\rule[-0.200pt]{12.286pt}{0.400pt}}
\put(630.0,272.0){\rule[-0.200pt]{0.400pt}{1.686pt}}
\multiput(556,629)(2.379,-20.619){3}{\usebox{\plotpoint}}
\multiput(562,577)(2.836,-20.561){3}{\usebox{\plotpoint}}
\multiput(570,519)(3.904,-20.385){2}{\usebox{\plotpoint}}
\multiput(579,472)(4.070,-20.352){2}{\usebox{\plotpoint}}
\multiput(587,432)(5.311,-20.064){2}{\usebox{\plotpoint}}
\put(599.82,384.16){\usebox{\plotpoint}}
\put(605.80,364.31){\usebox{\plotpoint}}
\put(612.49,344.67){\usebox{\plotpoint}}
\put(619.01,324.97){\usebox{\plotpoint}}
\put(625.61,305.35){\usebox{\plotpoint}}
\put(630.00,285.16){\usebox{\plotpoint}}
\put(630,279){\usebox{\plotpoint}}
\sbox{\plotpoint}{\rule[-0.400pt]{0.800pt}{0.800pt}}%
\put(463,123){\usebox{\plotpoint}}
\put(463.0,123.0){\rule[-0.400pt]{0.800pt}{121.895pt}}
\end{picture}
\caption{ \label{fig: Edge}
The stability of the quasi-liquid remaining on an edge 
as a function of the supersaturation relative to ice.
We plot the case of $W^{\rm S}(\delta)$. For $W^{\rm L}(\delta)$ , 
 qualitatively similar results are obtained.
The vertical solid line at supersaturation 0.167 
indicates the water saturation point.
The horizontal line at $\delta^{\rm S}_{\rm eq}$ indicates 
the equilibrium thickness of a quasi-liquid layer 
on a planar surface, given by Eq.(\ref{eq: Deqflat1}).
We used T$=-15^{\circ}{\rm C}$ $ (258.15 {\rm K})$,  
${\sigma}_{(10{\bar1}0)}=129.82 {\rm erg/cm^2}$ and 
$\Delta \sigma_{(\rm edge)}=36.2 {\rm erg/cm^2}$. 
We also used the experimental values at $-15^{\circ}{\rm C}$,  
$\sigma_{\rm s\ell(edge)}=24.25  {\rm erg/cm^2}$ and 
$\sigma_{\rm \ell v}=78.25 {\rm erg/cm^2}$ \cite{Cloud} 
(and therefore $\sigma_{\rm vs(edge)}=138.7 {\rm erg/cm^2}$).
This model is justified when $\delta $ is larger than the 
mono-layer thickness 
and $h$ is sufficiently large. 
Thus we ignore 
the unstable solutions existing for 
$\delta< 3 \stackrel{\circ}{\rm A}$ 
and $h< 30\stackrel{\circ}{\rm A}$. 
This type of figure can be obtained  
when ${\sigma}_{(10{\bar1}0)} > 128.9 {\rm erg/cm^2}$.
When $\mu_{\rm v}-\mu_{\rm s}\rightarrow \infty$, 
the right-hand side of Eq.(\ref{eq: curvcon}) 
converges to 133.07 ${\rm erg/ cm}^3$. 
This value is reasonable  
for a rough face $(10{\bar 1}0)$ having no quasi-liquid layer.
(c.f. $\sigma_{\rm vs(10{\bar 1}0)}=\rho Q_{\rm s}/(2 N_{\rm B})
=128 {\rm erg/cm^2}$, where $Q_{\rm s}=8.5\times 10^{-13}$erg/molecule ).
The critical supersaturation $\zeta(P_{\rm c})$ is $0.265$.
}
\end{center}
\end{figure*}

Numerical solutions of Eqs.(\ref{eq: Eqprism1}) and (\ref{eq: Eqprism2}) 
are plotted in Fig.\ref{fig: Edge}.  
(Here we have used Eqs.(\ref{eq: mup1})-(\ref{eq: mup3}) again.)
Remarkably, we find a stable equilibrium state when 
$P_{\rm v} < P_{\rm c}$.
In this case, a snow crystal with small $\delta$ and $h$ evolves 
toward a  stable state, at which point it ceases growing.

When $P_{\rm v}> P_{\rm c}$ $(>P_{\rm \ell})$, 
no stable solution satisfying Eqs.(\ref{eq: Eqprism1})
and (\ref{eq: Eqprism2}) is found. 
It is interesting that the thickness $\delta$ continues to grow in this case.
This suggests that a novel instability arises on the edges of snow
crystals when $P_{\rm v} > P_{\rm c}$.
Of course, in a real system, 
$\delta$ cannot continue to grow indefinitely, 
and thus this result reveals a limitation of our model. 
Our model breaks down 
when $\delta$ becomes comparable with $h$, 
and our simplification regarding the shapes of 
the quasi-liquids on the edges is no longer justified.
In a real system, as the thickness of the quasi-liquid grows, 
eventually it begins overflowing onto the neighboring 
faces $(10\bar{1}0)$ or $(0001)$ 
to reduce the surface energy.
We interpret the continuous growth of $\delta$ in our model 
as corresponding to the continuous overflowing of quasi-liquid 
in a real system. 
If the neighboring face is rough at the molecular level, 
the overflowing quasi-liquid sticks to the neighboring face, 
and immediately turns to be solid near the edge.   
As we show in the next section, this interpretation 
allows our model to explain the phase
diagram of snow crystals.

\section{Phase diagram}
\label{sec: 6}
In this section we give a new description of the snow phase diagram. 
Let us reconsider the discussion of  Section\ref{sec: 3}.
In the temperature regions 
(I), $0^{\circ}$C to $-4^{\circ}$C, and 
(IV), below $-22^{\circ}$C, 
the quasi-liquid on the edges and vertices does not play a special role.  
Thus we focus on the temperature regions (II) and (III), $-4^{\circ}$C to
$-22^{\circ}$.

In the temperature region (II), $-4^{\circ}$C to $-10^{\circ}$C, 
the six prism faces $(10\bar{1}0)$, 
all the edges and all the vertices are  covered with quasi-liquid, 
while the two basal faces $(0001)$ are rough at the molecular level 
and have no quasi-liquid layers.
As seen in Fig. \ref{fig: Edge},
when the vapor pressure $P_{\rm v}$ is low, 
the quasi-liquids on the edges and vertices are stable.
In this case, the edges and the vertices 
do not play a special role in the snow crystal formation.   
However, when $P_{\rm v}$ is higher than a certain value 
$P_{\rm c}$ $(>P_{\ell})$, 
the quasi-liquid on the edges and  vertices overflows 
onto the neighboring faces.
The quasi-liquid overflowing onto $(10\bar{1}0)$ 
spreads over the quasi-liquid layer, 
while 
the quasi-liquid overflowing onto rough $(0001)$ sticks 
near the edges and vertices and turns to solid.
In this case, two types of overflow are possible.
One is overflow from the vertices, and 
the other is  overflow from the edges where the basal and  prism faces meet.
If the former becomes the main type of overflow, 
the resultant snow crystal should grow like a needle.
If the latter becomes the main overflow according to our model, 
the resultant snow crystal should grow like a sheath.

In the temperature region (III), $-10^{\circ}$C to $-22^{\circ}$C, 
the edges and the vertices are covered with  quasi-liquid, 
while the basal $(0001)$ and  prism  $(10{\bar 1}0)$ faces 
possess no quasi-liquid layer. 
The basal faces are smooth 
and the prism faces are rough at the molecular level.  
Here again, as seen in Fig. \ref{fig: Edge}, when $P_{\rm v}$ is low, 
the quasi-liquid is stable, 
and thus the snow crystal grows according to the theory of Kuroda and Lacmann.
When $P_{\rm v}$ is higher than 
$P_{\rm c}$ $(>P_{\ell})$, 
the quasi-liquid overflows onto the neighboring faces.
The quasi-liquid overflowing on to smooth $(0001)$ 
spreads over $(0001)$ and turns to solid, 
while 
the quasi-liquid overflowing onto rough $(10\bar{1}0)$
sticks near the edges and vertices and turns to solid. 
In this case, three types of overflow are possible.
The first is  overflow from the vertices, 
the second is overflow from the edges 
where the basal and the prism faces meet, 
and the third is overflow from the edges where two prism faces meet.
If the amounts of quasi-liquid involved in all overflows are of 
the same order,  
the resultant snow crystal should grow like a sector at about $-10^{\circ}$C.
If the first and (or) the third overflow are the main types of overflow, 
the resultant snow crystal should  
grow like a dendrite at about $-15^{\circ}$C. 
Contrastingly if the first and (or) 
the third overflow are the main overflows, 
but the surface structure of prism face changes   
from a rough surface to a smooth surface with temperature falling,     
the resultant snow crystal should grow like a sector again.
This is because the continuous overflow 
spreads over smooth prism faces and 
turns to solid successively.  
In this case, the overflow can act as a step source for prism faces 
and thus the snow crystal grows in the manner hypothesized by 
Frank  \cite{Frank} and found 
in the simulation of Yokoyama and Kuroda \cite{YK}.
We believe that this type of sector can be observed at about $-22^{\circ}$C.  

We note that the critical supersaturation shown in Fig.\ref{fig: Edge}
 is 0.265  at $-15^{\circ}$C, 
which agrees with the experimental supersaturation, 
for which dendrites are observed above 0.20 
(see Fig.\ref{fig: snowdiagram}).

\section{Summary}

In this paper, we have shown theoretically that 
quasi-liquid layers exist on the edges and vertices of snow crystals 
between $-4^{\circ}$C and $-22^{\circ}$C, 
for which temperatures there is no quasi-liquid 
on the faces $(0001)$ and  $(10{\bar 1}0)$.

To investigate the curvature effect on quasi-liquid layers 
at the edges and  vertices, 
we considered an ice particle coated with a quasi-liquid layer, 
and derived the equilibrium over a range of supersaturation values.
We found that below the water saturation point, 
the equilibrium state is uniquely determined. 
However, above this point, 
another equilibrium state exists 
because it is possible for bulk supercooled water to exist. 
Above the critical supersaturation point, 
no equilibrium state exists.  
We showed that 
the quasi-liquid layers for equilibrium states of a particle   
are thicker than the quasi-liquid layer 
for the equilibrium state of a planar  surface.

Next, we  considered a two-dimensional crystal. 
We found that there are temperatures for which the six prism faces 
possess no quasi-liquid, while the edges and vertices are 
covered with quasi-liquid.
We have examined the stability of this system.  
We found that below the critical saturation, 
which is higher than the 
water saturation,
the quasi-liquid layer exists in a stable state on the edges. 
However, above the critical supersaturation,  
the quasi-liquid becomes unstable, and continues to grow indefinitely.
We interpret this indefinite growth as 
implying overflow onto the neighboring faces in a real system.

On the basis of our results, 
we have proposed a new description of snow crystal growth 
according to which 
the unstable growth of snow crystals, 
which prefer the edges and the vertices, is due to 
the overflow of the quasi-liquid 
from the edges and the vertices onto neighboring 
faces that are rough and have no quasi-liquid layers. 
We have shown that 
this overflowing occurs above the water saturation point 
and that these surface conditions of snow crystals 
 are realized between $-4^{\circ}$C and $-22^{\circ}$C. 
This description naturally accounts 
the relation between the morphological
instability and the water saturation in the snow phase diagram.

While we have succeeded in explaining 
the snow phase diagram qualitatively, for a more complete understanding, 
the details of crystal growth  
should be clarified by three-dimensional simulation.
We also would like to show that secondary branches of the dendrites are 
produced by the overflow of quasi-liquid 
from the edges and vertices of the primary branches.
These are future problems. 

In this paper, we have focused on the quasi-liquid of snow crystals. 
However surface melting is observed 
in many classes of solids, including metals, 
semiconductors, solid rare gases \cite{DasFuWet}.
The model and method presented in this paper are also applicable to these
materials, and should be useful for the investigation of general properties of
 surface melting. 

\begin{acknowledgments}
Special thanks are due to Professor 
Y. Furukawa and M. Sato 
for many fruitful discussions. 
\end{acknowledgments}

\end {document}